\input harvmac
\def\Title#1#2#3{#3\hfill\break \vskip -0.35in
\rightline{#1}\ifx\answ\bigans\vskip.2in
\else\pageno1\vskip.2in\fi \centerline{\titlefont #2}\vskip .1in}


\def\bra#1{\langle #1 |}
\def\ket#1{| #1\rangle}
\def\braket#1#2{\langle \, #1 \, | \, #2 \, \rangle}

\def\S{{\cal{S}}}

\def\a{\alpha}
\def\b{\beta}
\def\g{\gamma}
\def\eps{\epsilon}
\def\D{{\rm{D}}}
\def\UP{{\cal UP}}

\def\R{\hbox{\rm I \kern-5pt R}}

\font\ticp=cmcsc10
\def\ajou#1&#2(#3){\ \sl#1\bf#2\rm(19#3)}
%
%
\lref\griff{R.B.~Griffiths, \ajou J. Stat. Phys. &36 (84) 219.}

\lref\hartleone{J.B.~Hartle, in {\it Quantum Cosmology and Baby
Universes}, Proceedings of the 1989 Jerusalem Winter School on
Theoretical Physics,
ed.~by S.~Coleman, J.~Hartle, T.~Piran, and S.~Weinberg,
World Scientific, Singapore (1991).}

\lref\grifflogic{R.B.~Griffiths, \ajou Found. Phys. &23 (93) 1601.}

\lref\griffithschqr{R.B.~Griffiths, \ajou Phys. Rev. &A54 (96) 2759.}

\lref\omnes{R.~Omn\`es, \ajou J. Stat. Phys. &53 (88) 893.}

\lref\omnespla{R.~Omn\`es, \ajou Phys. Lett. A &187 (94) 26.}

\lref\sorkin{R.D.~Sorkin, \ajou Mod. Phys. Lett. &A9 (94) 3119.}

\lref\kentscripta{A.~Kent, 
in {\it Modern Studies of Basic Quantum Concepts and Phenomena}, 
Proceedings of the 104th Nobel Symposium, Gimo, June 1997, 
Physica Scripta T76 78-84 (1998).} 

\lref\bohm{D.~Bohm, \ajou Phys. Rev. &85 (52) 166.}

\lref\grw{G.~Ghirardi, A.~Rimini and T.~Weber, \ajou Phys. Rev. &D34
(86) 470.}

\lref\av{Y.~Aharonov and L.~Vaidman, \ajou J. Phys. A &24 (91) 2315.} 

\lref\omnesbook{R.~Omn\`es, {\it The Interpretation of Quantum
Mechanics}, Princeton University Press, Princeton (1994).}

\lref\Ishamql{C.J.~Isham, \ajou J. Math. Phys. &23 (94) 2157.}

\lref\ILqtl {C.J.~Isham and N.~Linden, \ajou J. Math. Phys. &35 (94) 5452.}

\lref\ILSclass{ C.J.~Isham, N.~Linden and S.~Schreckenberg, 
\ajou J. Math. Phys. &35 (94) 6360.}

\lref\brunhalliwell{T.~Brun and 
J.~Halliwell, \ajou Phys. Rev. &D54 (96) 2899.}

\lref\cohen{O.~Cohen, \ajou Phys. Rev. &A51 (95) 4373.}

\lref\dowkerhalliwell{H.F.~Dowker and J.~Halliwell, \ajou
Phys. Rev. &D46 (92) 1580.} 

\lref\rudolphtwo{O.~Rudolph, \ajou J. Math. Phys. &37 (96) 5368.}

\lref\rudolphone{O.~Rudolph, \ajou Int. J. Theor. Phys. &35 (96) 1581.}

\lref\gmhsantafe{M.~Gell-Mann and J.B.~Hartle in {\it Complexity, Entropy,
and the Physics of Information, SFI Studies in the Sciences of
Complexity}, Vol.
VIII, W.~Zurek (ed.),  Addison Wesley, Reading (1990).}

\lref\gmhtimeasymmetry{M.~Gell-Mann and J.B.~Hartle in 
{\it Physical Origins of Time Asymmetry}, J.~Halliwell,
J.~P\'erez-Mercader and W.~Zurek (eds.), Cambridge University Press, 
Cambridge (1994).}

\lref\gmhprd{M.~Gell-Mann and J.B.~Hartle, \ajou Phys. Rev. 
&D47 (93) 3345.}

\lref\gmhstrongd{M.~Gell-Mann and J.B.~Hartle, 
{U}niversity of California, Santa   Barbara preprint UCSBTH-95-28;
gr-qc/9509054.}

\lref\gmhequiv{M.~Gell-Mann and J.B.~Hartle, 
{U}niversity of California, Santa Barbara preprint UCSBTH-94-09;
gr-qc/9404013.}

\lref\gellmannbook{M.~Gell-Mann, {\it The Quark and the Jaguar}, 
Little, Brown and Co, London (1994).}

\lref\lindenpriv{N.~Linden, private communication.}

\lref\dowkerkentone{F.~Dowker and A.~Kent, 
\ajou J. Stat. Phys. &82 (96) 1575.}

\lref\dowkerkenttwo{F.~Dowker and A.~Kent, \ajou Phys. Rev. Lett. &75
(95) 3038.}

\lref\kentquasi{A.~Kent, \ajou Phys. Rev. &A54 (96) 4670.}

\lref\kentcontra{A.~Kent, \ajou Phys. Rev. Lett. &78 (97) 2874.}

\lref\ghcomment{R.~Griffiths and J.~Hartle, \ajou Phys. Rev. Lett. &81
(98) 1981.} 

\lref\kentreply{A.~Kent, \ajou Phys. Rev. Lett. &81 (98) 1982.} 

\lref\kentmcelwaine{A.~Kent and J.~McElwaine, \ajou Phys. Rev. &A55
 (97) 1703.} 

\lref\kentbohmbook{A.~Kent, in 
{\it Bohmian Mechanics and Quantum Theory: An Appraisal}, edited
  by J. Cushing, A. Fine, and S. Goldstein (Kluwer Academic Press, Dordrecht,
  1996).}

\lref\mcelwainemaxinfo{J.~McElwaine, \ajou Phys. Rev. &A56 (97) 1756.} 

\lref\pazzurek{J.~Paz and W.~Zurek, \ajou Phys. Rev. &D48 (93) 2728.}

\lref\giardinarimini{I.~Giardina and A.~Rimini, \ajou Found. Phys. &26
(96) 973.}

\lref\goldsteinpage{S.~Goldstein and D.~Page, \ajou Phys. Rev. Lett. 
&74 (95) 3715.}

\lref\bgl{P.~Busch, M.~Grabowski and P.~Lahti, {\it Operational Quantum
Physics}, Lecture Notes in Physics m31, Springer-Verlag, 
Berlin (1995).} 

\Title{\vbox{\baselineskip12pt\hbox{ }\hbox{}{}
}}
{\vbox{\centerline {Quantum Histories and Their Implications}}}{~}

\centerline{{\ticp Adrian Kent\foot{Email: a.p.a.kent@damtp.cam.ac.uk}}}
\vskip.1in
\centerline{\sl Department of Applied Mathematics and
Theoretical Physics,}
\centerline{\sl University of Cambridge,}
\centerline{\sl Silver Street, Cambridge CB3 9EW, U.K.}

\bigskip
\baselineskip24pt

\centerline{\bf Abstract} { Classical mechanics and standard
Copenhagen quantum mechanics respect subspace implications.
For example, if a particle is confined in a particular
region $R$ of space, then in these theories we can deduce that
it is confined in regions containing $R$.  However, subspace 
implications are generally violated by versions of quantum 
theory that assign probabilities to histories, such as 
the consistent histories approach.  

I define here a new criterion, ordered consistency, which refines
the criterion of consistency and has the property that 
inferences made by ordered consistent sets do not 
violate subspace relations.  This raises  
the question: do the operators defining our 
observations form an ordered consistent history?  If so, 
ordered consistency defines a version of 
quantum theory with greater predictive power than 
the consistent histories formalism.  
If not, and our observations are defined by a non-ordered consistent quantum
history, then subspace implications are not generally valid.  
\vskip 10pt {\it PACS:} 03.65.Bz\hfill\break }
\vfill\eject 

\newsec{Introduction}

We take it for granted that we can infer quantitatively less precise 
statements from our observations.
For example, if we know that an atom 
is confined in some region $R$ of space, we believe we are 
free to assume for calculational purposes 
only that it lies in some larger region containing $R$. 
Our understanding of the world and our interpretation of everyday 
experience tacitly rely on subspace implications of this general
type: if a physical quantity is known to lie within a range $R_1$, 
then it lies in all ranges $R_2 \supset R_1$. 

In classical mechanics, subspace inferences follow from the 
correspondence between physical states and points in phase space: 
if the state of a system lies in a 
subset $S_1$ of phase space, it lies in all subsets $S_2 \supset S_1$.
Similarly, in Copenhagen quantum theory, they hold since if the 
state of a quantum system lies in a subspace
$H_1$ of Hilbert space, it lies in all subspaces $H_2 \supset H_1$. 

However, neither classical mechanics nor (presumably) 
Copenhagen quantum theory is fundamentally correct.   
If the basic principles of quantum theory apply to the universe
as a whole, then a post-Copenhagen interpretation of 
quantum theory seems to be needed, and any justification of 
the subspace implications must ultimately be given in terms
of that interpretation. 
Though it may seem hard to imagine how to make sense of nature
without allowing subspace inferences, there are versions 
of quantum theory in which they do not hold. 
In particular, this is true of recent  
attempts to develop history-based formulations of 
quantum theory,\refs{\griff, \grifflogic, \omnes, \gmhsantafe, 
\Ishamql} which rely on the notion of consistent or
decoherent sets of histories.  

This paper suggests a way in which quantum theory can plausibly
be interpreted via statements about histories, without violating
subspace implications --- the motivation being that both history-based
interpretations and subspace implications seem natural.  
The interpretation which results, a refinement of the consistent 
histories approach based on ordered consistent sets
of histories, is certainly not the ultimate answer to the problems
of quantum theory.  It does not solve the measurement
problem, for example.  But perhaps it is a step in the right direction, or at
least in an interesting direction.  It also helps to make precise
the question as to what it would mean if subspace implications 
actually {\it were} violated in nature, which the last part 
of the paper examines.  

The language of quantum histories
may not necessarily be the right way to interpret the quantum theory of 
closed systems.  Bohm theory\refs{\bohm} and dynamical collapse 
theories\refs{\grw} are particularly interesting alternatives,
for example.  Quantum histories still play a r\^ole in these
theories as usually interpreted, but it is certainly possible 
to interpret them so as to ascribe no ontological status
to the evolving quantum state, in which case quantum histories have no
ontological status either, and the orderings amongst them become 
physically irrelevant.  

In this paper, which focusses on the relation between quantum 
histories and orderings, I take it for granted --- as an
interesting hypothesis, not a dogma --- that quantum
histories have some ontological status.  
I will focus on the consistent histories approach. 
This is not to suggest that there are no other interesting
quantum histories approaches.  Other ideas are outlined in
Ref. \refs{\sorkin}, Section VII
of Ref. \refs{\dowkerkentone}, and Ref. \refs{\kentscripta}, for 
example.  But the discussion of this paper is best carried out
within some specific formulation quantum theory, and consistent histories 
is the most widely studied example.  
As an approach to closed system 
quantum histories, it has many interesting features.
Its main drawback is that it does not solve the measurement problem, 
which in the language of consistent histories takes the form of
a set selection problem.  The search for ways to refine the definition
of consistency in order to solve, or at least reduce, the set
selection problem is an independent motivation for the definition
of ordered consistency proposed below. 

\newsec{Partial Ordering of Quantum Histories} 

We consider versions of quantum theory that
assign probabilities to {\it histories} --- i.e., to collections 
of events.  
There are several reasonably standard ways of representing an
event in quantum theory.  The simplest is as a projection 
operator, labelled by a particular time,
corresponding to a statement about an observable in non-relativistic 
quantum mechanics.  Thus, the statement
that a particle was in the interval $I$ at time $t$
is represented by the projection 
\eqn\interval{P_I =  \int_{x \in I} \ket{x} \bra{x} d^3 x   \, .}
The representation of events as projections can be generalised
to {\it quantum 
effects}\refs{\rudolphone, \rudolphtwo, \kentscripta}, 
defined by operators $A$ such that $A$ and $I-A$ are both 
positive.\foot{Different definitions can be found: this, 
the simplest, is adequate for our purposes.} 
Events can also be defined, at least formally,
in the path integral version of quantum theory by 
dividing the set of paths into exclusive subsets.  For example, by 
considering the appropriate integrals one can attach probabilities to 
the events that a particle did, or did not, enter a particular 
region of space-time\refs{\hartleone}. 
Further generalisations have been discussed by 
Isham, Linden and collaborators.\refs{\Ishamql,\ILqtl,\ILSclass}
 
Each of these 
representations has a natural partial ordering.
For projections, we take $A \geq B$ if 
the range of $B$ is a subspace of that of $A$.  
This corresponds to the logical implication
of Copenhagen quantum theory already mentioned: if the state of a 
quantum system lies in the range of $B$, then it necessarily 
lies in the range of $A$. 
For quantum effects, we take $A \geq B$ if and only if 
$(A-B)$ is positive.  
For events defined by the position space path 
integral, we take the statement that a particle entered $R_1$ to imply
the statement that it entered $R_2$ if $R_1 \subseteq R_2$. 

We define a quantum history as a collection of quantum events 
and extend the partial ordering to histories in the natural way.  
For example, if we take quantum histories in non-relativistic
quantum mechanics to be defined by sequences of projections
at different fixed times, we compare 
two quantum histories $H = \{ A_1 , t_1 ; \ldots ; 
A_n ; t_n \}$ and $H' = \{ A'_1 , t'_1 ; \ldots ; A^{'}_{n'} , t^{'}_{n'} \}$ 
as follows.  First add to each history the identity operator at 
every time at which it contains no proposition and the other does, 
so obtaining relabelled representations of the 
histories as $H = \{ A_1 , T_1 ; \ldots ; 
A_N ; T_N \}$ and $H' = \{ A'_1 , T_1 ; \ldots ; A'_N , T_N \}$.
Then define the partial ordering by 
taking $H \geq H'$ if and only if $A_i \geq A'_i$ for all $i$ 
and $H > H'$ if $H \geq H'$ and $H \neq H'$.
This ordering was first considered in the consistent histories
literature by Isham and Linden\refs{\ILqtl}, whose work, and
its relation to the ideas of this paper, are discussed in
the Appendix.  

\newsec{Consistent Histories} 

The consistent sets of histories for a closed quantum 
system are defined in terms of the space of states $\cal{H}$, the initial 
density matrix $\rho$, and the hamiltonian $H$.  
In the simplest version of the consistent histories 
formulation of non-relativistic quantum mechanics, sets of 
histories correspond to sets of projective decompositions.
In order to be able to give a physical interpretation of any of
the consistent sets, we need also to assume that standard observables, 
such as position, momentum and spin, are given. 
Individual quantum events 
are defined by members of projective decompositions of the identity
into orthogonal 
hermitian projections $\sigma = \{ P^{i} \}$, with
\eqn\conds{\sum_i P^{i}  = 1 \quad {\rm{and}}\quad
P^{i} P^{j} = \delta_{ij} P^{i} \, .}
Each such decomposition defines a complete and exclusive list of 
events at some fixed time, and a time label is thus generally attached
to the decompositions and the projections: the labels are omitted here,  
since the properties of interest here depend only on the time ordering
of events. 

Suppose now we have a set of 
decompositions $\S = \{ \sigma_1 , \ldots , \sigma_n \}$.  
Then the histories given by choosing one projection from each of 
the decompositions $\sigma_j$ in all possible ways 
define an exhaustive and exclusive set of alternatives. 
We follow Gell-Mann and Hartle's definitions, and
say that $\S$ is a {\it consistent} (or {\it medium decoherent})
set of histories if
\eqn\decohgmha{ \Tr (  P^{i_n}_n \ldots P^{i_1}_1
                               \rho P^{j_1}_1
\ldots P^{j_n}_n ) =
\delta_{i_1 j_1 } \ldots \delta_{i_n j_n } p(i_1 \ldots i_n ) \, .}
When $\S$ is consistent, $p(i_1 \ldots i_n )$ is the probability of the
history $H = \{ P^{i_1}_1 , \ldots , P^{i_n}_n \}$.
We will want later to discuss the properties of individual histories 
without reference to any fixed consistent set, and we 
define a {\it consistent history} to be a history which 
belongs to some consistent set $\S$. 
Finally, we say the set
\eqn\extend{
\S' =  \{\sigma_1,\dots, \sigma_k, \tau, \sigma_{k+1},\dots,
\sigma_n\}}
is a {\it consistent extension} of a consistent set of histories
$\S =  \{\sigma_1 , \ldots , \sigma_n \}$ by 
the set of projections $\tau = \{Q^i : i = 1, \dots, m\}$ if $\tau$
is a projective decomposition and $\S'$ is consistent.  

Suppose now that we have a collection of data defined by the history 
\eqn\history{H = \{ P^{i_1}_1 , \ldots , P^{i_n}_n \}}
which has non-zero
probability and belongs to the consistent set $\S$. 
This history might, for example, describe the results of a series
of experiments or the observations made by an
observer.  Given a choice of consistent 
extension $\S'$ of $\S$, we can make probabilistic 
inferences conditioned on the history $H$. 
For example, if $\S'$ has the above form, the histories
extending $H$ in $\S'$ are
$H^i = \{ P^{i_1}_1 , \ldots , 
P^{i_k}_k , Q^i , P^{i_{k+1}}_{k+1}, \ldots , P^{i_n}_n \}$
and the proposition $Q^i$ has conditional probability 
\eqn\conditional{p( Q^i | H ) = p( H^i ) / p(H) \, .}
These probabilities and conditional probabilities are
the same in every consistent set which includes $H^i$ and (hence)
$H$.  However, when we want to emphasize that the calculation can 
be carried out in some particular set $\S$, we will attach a suffix.
For example, we might write 
$p_{\S'} ( Q^i | H ) = p_{\S'} ( H^i ) / p_{\S} (H)$ for 
the above equation. 

The formalism itself gives no way of choosing any particular consistent
extension.  In the view of the original developers of the consistent 
histories approach, the 
different $\S'$ are to be thought of as ways of producing 
possible pictures of the past and future 
physics of the system which, though generally
incompatible, are all equally valid.  More formally, they can be 
seen as incompatible logical structures which allow different 
classes of inferences from the data\refs{\omnesbook,\griffithschqr}.

It is this freedom in the choice of consistent extension 
which, it has been argued elsewhere\refs{\dowkerkentone, \kentquasi,
\kentcontra,  \dowkerkenttwo, \pazzurek, \giardinarimini,
\kentbohmbook, \kentscripta}
gives rise to the most serious problems 
in the consistent histories approach.  
Standard probabilistic predictions and deterministic retrodictions
can be reproduced in the consistent histories formalism by making
an ad hoc choice of consistent set, but cannot be derived from
the formalism itself.    
In fact, it is almost never possible to make any unambiguous 
predictions or retrodictions: there are almost always an 
infinite number of incompatible consistent extensions of the set containing
a given history dataset\refs{\dowkerkentone, \dowkerkenttwo}. 

The problem is not simply that the formalism supplies descriptions
of physics which are complementary in the standard sense, 
although that in itself is sufficient to ensure that the
formalism is only very weakly predictive.  
Even on the assumption that we will continue to observe quasiclassical
physics, no known interpretation of the formalism allows us to 
derive the predictions of classical mechanics and Copenhagen
quantum theory.\refs{\kentquasi}.  
Hence the set
selection problem: probabilistic predictions can only be made
conditional on a choice of consistent set, yet the consistent
histories formalism gives no way of singling out any particular set or
sets as physically interesting.

One possible solution to the set selection problem would be an axiom
which identifies a unique physically interesting set, or perhaps a
class of such sets, from the initial state and the dynamics.  Another
would be the identification of a physically natural measure on the
space of consistent sets, according to which the physically relevant
consistent set is randomly chosen.  
Possible set selection criteria have been 
investigated\refs{\omnespla, \kentmcelwaine}, 
but no generally workable criterion has emerged.\foot{
With different motivations, Gell-Mann and Hartle have explored  
a ``strong decoherence'' criterion which is intended to reduce the
number of consistent sets.\refs{\gmhstrongd} 
However, under their present definition, every consistent set 
is strongly decoherent.\refs{\kentscripta}}  

\newsec{Consistent Sets and Contrary Inferences: A Brief Review}

A further reason for believing that the consistent histories formalism
is at best incomplete comes from considering the logical relations 
among events in different consistent sets. 
We say that two projection operators $P$ and $Q$ are {\it complementary}
if they do not commute: $PQ \neq QP$. 
We say that they are {\it contradictory} if they sum to the identity, 
so that $ P = 1 - Q$ and $PQ = QP = 0$, and that they 
are {\it contrary} if they are orthogonal and not contradictory, 
so that $PQ = QP = 0$ and $ P < 1 - Q$.  

Contradictory inferences are never possible in the consistent
histories formalism, but it is easy to find examples of contrary 
inferences from the same data.\refs{\kentcontra}
For instance, consider a quantum system whose hamiltonian is zero and
whose Hilbert space $\cal{H}$ has dimension greater than two, 
prepared in the state $\ket{a}$. 
Suppose that the system is left undisturbed from time $0$ until time
$t$, when it is observed in the state $\ket{c}$, where 
$0 < \, | \braket{a}{c} | \, \leq 1/3$: i.e. a single quantum 
measurement is made, and the outcome probability is 
less than $1/9$.
In consistent histories language, we have initial density matrix 
$\rho = \ket{a} \bra{a}$ and the single datum corresponding
to the history 
$H = \{ P_c \}$ from the 
consistent set $\S = \{ \{P_c , 1 - P_c \} \}$, 
where the projection $P_c = \ket{ c} \bra{c}$ is taken at time $t$. 

Now consider consistent extensions of $\S$ of the form 
$\S_b = \{ \{P_b , 1 - P_b \}, \{P_c , 1 - P_c \} \}$, 
where $P_b = \ket{ b} \bra{b}$ for some 
normalised vector $\ket{b}$ with the property
that 
\eqn\vone{ \braket{c}{b} \, \braket{b}{a} \, = \, 
\braket{c}{a} \, . } 
It is easy to verify that $\S_b$ is consistent and that 
the conditional probability of $P_b$ given $H$ is $1$.  
It is also easy to see that there are
at least two mutually orthogonal vectors $\ket{b}$ 
satisfying \vone .  For example, let $\ket{v_1} , \ket{v_2} ,
\ket{v_3}$ 
be orthonormal
vectors and take $\ket{a} = \ket{v_1}$ 
and $\ket{c} = \lambda \ket{v_1} + \mu \ket{v_2} $, where 
$ | \lambda |^2 + | \mu |^2 = 1$.  
Then the vectors 
\eqn\vecs{\ket{b_{\pm}} = ( |\lambda |^2 + { |\mu |^2 \over x} )^{-1/2} (
\lambda \ket{v_1} + {{\mu} \over {x}} \ket{v_2} \pm 
{{ (x-1)^{1/2} \mu} \over {x}} \ket{v_3}) }
both satisfy \vone\ and are orthogonal if $x$ is real and
$x^2 | \lambda|^2 = (x-2) ( 1 - |\lambda |^2 )$, which has solutions for 
$| \lambda | \leq 1/3$. 
Thus the consistent sets $\S_{b_{\pm}}$ give contrary 
probability one retrodictions. 

Some brief historical remarks are in order. 
The existence of contrary inferences in the consistent histories 
formalism, though easy to show, was noticed only quite recently.
In particular, it
was not known to the formalism's original developers.\foot{I am 
grateful to Bob Griffiths, Jim Hartle, and Roland Omn\`es for helpful 
correspondence on this point.}
It was first explicitly pointed out,
and its implications for the consistent 
histories formalism were first examined, in Ref. \refs{\kentcontra}.
Further discussion can be found in Refs. \refs{\ghcomment, \kentreply}.  
However, a noteworthy earlier consistent histories analysis 
of an example in which contrary inferences arise can be found 
in a critique by Cohen\refs{\cohen} of 
Aharonov and Vaidman's interpretation\refs{\av}
of one of their intriguing examples of pre- and 
post-selection.\foot{I am grateful to Oliver Cohen and Lucien Hardy
for drawing this reference to my attention.} 
As noted in Ref. \refs{\kentcontra}, Cohen's analysis 
miscontrues the consistency criterion: however, this
error does not affect its derivation of contrary inferences.

Now, the existence of contrary inferences in the consistent 
histories formalism needs to be interpreted with care.  
It is {\it not} true that, in any given consistent set, 
two different contrary propositions can be inferred with probability
one.  The inferences made within any given consistent set lead
to no contradiction.  The picture of physics given by any given
consistent set may or may not be considered natural or 
plausible --- depending on one's intuition and the criteria one
uses for naturality --- but it is 
not logically self-contradictory.  
It {\it is} however true, as a mathematical statement about the 
properties of the consistent histories formalism, that the 
propositions inferred in the two different sets correspond
to contrary projections.  
The formalism makes no physical distinction among different consistent
sets, and so requires us to conclude that two equally valid pictures 
of physics can be given, in which contrary events take place.  

To put it more formally, the consistent histories approach 
can be interpreted as setting out rules of reasoning according 
to which, although physics 
can be described by any of infinitely many equally valid pictures, 
only one of those pictures may be considered in any given argument. 
Such an interpretation ensures --- tautologically --- that 
no logical contradiction arises, 
even when the pictures contain contrary inferences.  

The consistent histories formalism, in other words, 
gives a set of rules for producing possible pictures of 
physics within quantum theory, and these rules themselves 
lead to no logical inconsistency.  
However, consistent historians claim much more, arguing that the 
formalism defines a natural and scientifically unproblematic 
interpretation of quantum theory.  
Indeed, the consistent histories literature tends
to suggest that the descriptions of 
physics given by consistent historians are simply and evidently 
the correct descriptions which emerge from quantum theory, 
so that, in querying them, one necessarily queries quantum 
theory itself.\foot{For example, the consistent histories 
interpretation of quantum mechanics has been referred to as
``the interpretation of quantum mechanics''\refs{\omnesbook} 
and even as simply ``quantum mechanics''\refs{\gmhequiv}.}   

This seems patently false.  
The most basic premise of consistent historians --- 
that quantum theory is correctly interpreted by some sort of 
many-picture scheme --- leads to such trouble in explaining 
which particular picture we see, and why, that cautious scepticism seems 
only appropriate. 
Even if the premise were accepted, it would be essential to ask, of any 
particular many-picture scheme, whether its assumptions are 
natural and whether the descriptions of nature it produces are 
physically plausible or scientifically useful.  
The particular equations used to define consistent sets are, 
after all, simply interesting guesses: there is 
no compelling theoretical justification for them, and indeed, 
several different definitions of consistency have been 
proposed\refs{\griff,\gmhsantafe,\gmhprd}. 

My own view is that there are a number of compelling reasons 
for regarding the consistent histories interpretation, as it 
is presently understood, as scientifically unsatisfactory.  
However, as these questions have 
been explored in some detail elsewhere\refs{\dowkerkentone, \kentquasi, 
\kentcontra, \ghcomment, \kentreply, 
\dowkerkenttwo - \kentbohmbook, \kentscripta}, I here  
comment only on two specific problems raised by contrary inferences.    

First, the fact that we are to take as equally valid and correct
pictures of physics which include contrary inferences goes against
many well developed intuitions.  
No argument based on intuition alone can be conclusive,
but I think it must be granted that this one has some force.  
What use, it may reasonably be asked, is there in saying that 
in one picture of reality a particle genuinely went, with 
probability one, through slit A, and that in another picture the 
particle went, also with probability one, through the disjoint slit B?  Why 
should we take either picture seriously, given the other?  

On this point, it is worth noting that one of the advertised merits
of the consistent histories formalism\refs{\griff, 
\omnesbook, \griffithschqr}, is that it, unlike the 
Copenhagen interpretation, accommodates some (arguably) plausible 
intuitions about the behaviour of microsystems in between  
observations.  For example, the formalism allows us to say --- albeit
only as one of an infinite number of incompatible descriptions ---
that a particle observed at a particular detector was travelling
towards that detector before the observation, and that a particle
measured to have spin component $\sigma_x = s_x$ had that spin before the
measurement took place.  
As Griffiths and Omn\`es note,\refs{\griff, \omnesbook} informal 
discussions of experiments are often framed in terms which, if taken
literally, suggest that we can make this sort of statement 
about microsystems before a measurement is carried out.  
(``Was the beam correctly aligned going into the second
interferometer?'', or ``Do you think something crazy in the
electronics might have triggered [detector] number 3 just before
the particle got there?''\refs{\griff}, for example.)
Their intuition is that a good interpretation of quantum theory 
ought to give a way of allowing us to take such 
statements literally --- a criterion which, they suggest,  
the consistent histories approach satisfies. 

The intuition is, of course, controversial.  
A counter-intuition, which most interpretations of quantum theory
support, is that any description of a microsystem before a measurement
is carried out should be independent of the result of that
measurement. 

In any case, to the extent that any intuition is offered 
as a justification of the formalism, it seems reasonable 
to consider the fact that the formalism violates 
other strongly held intuitions.
Few experimenters, after all, can ever have intuitively concluded that 
the {\it entire} flux of their beam can sensibly be 
thought of as having followed any of several macroscopically distinct 
paths through the apparatus.  Yet this is what the above example, 
translated into an interference experiment, implies.  

The second, and probably deeper, problem is
that it seems very hard to justify the distinction, which consistent
historians are forced to draw, between contradictory 
inferences, which are regarded as {\it a priori} unacceptable, 
and contrary inferences, which are regarded as unproblematic.
Some justification seems called for, since the distinction is not an 
accidental feature: it is not that the 
formalism, for unrelated reasons, simply happens to exclude one 
type of inference and include the other.  The definition
of consistency is motivated precisely by the notion that, when two 
different sets allow a calculation of the probability of the same event  
(belonging, in the simplest case, to a single history in one set
and a combination of two histories in the other), the calculations
should agree.  This requires in particular that 
contradictory propositions $P$ and $(1-P)$ can never be 
inferred, since if the probability of $P$ is one in 
any set, it must be one in all sets, and so the probability of 
$(1-P)$ must be zero in all sets.
  
Now this last requirement is not absolutely essential, 
sensible though it may seem.  No logical 
contradiction arises in an interpretation 
of quantum theory which follows the basic interpretational ideas of 
the consistent histories formalism but which accepts all complete sets
of disjoint quantum histories, whether consistent or not, as defining
valid pictures of physics\refs{\dowkerkentone, \kentscripta}. 
In this ``inconsistent histories''
interpretation, contradictory inferences can generally be made 
by using different pictures.  This possibility is excluded by
a deliberate theoretical choice. 

It seems natural, then, having made this choice, 
to look for ways in which contrary inferences can similarly be
excluded.  This is the line of thought pursued
below.  Note, however, that the problem
of contrary inferences is not the only motivation for the ideas 
introduced below. 
Whatever one's view of the consistent histories formalism, it is interesting  
that an alternative formalism can be defined relatively simply.
It seems fruitful to ask which, if either, is to be preferred, and why.  
And, as we will see, ordered consistent sets raise independently interesting
questions about the quasiclassical world we actually observe. 

\newsec{Relation of Contrary Inferences and Subspace Implications}  

A contrary inference arises 
when there exist two consistent sets, $\S_1$ and $\S_2$,
both containing a history $H$, with the property that 
there are orthogonal propositions $P_1$ and $P_2$ which
are implied by $H$ in the respective sets, so that --- temporarily adding 
set suffices for clarity --- we have 
\eqn\contrary{ p_{\S_1} ( P_1 | H ) = p_{\S_2} ( P_2 | H ) = 1 \, .}
Now $p_{\S_2} ( (1 - P_2 ) | H ) = 0$, and since the probabilities
are set-independent and $p_{\S_1} ( P_1 | H )$ is nonzero, we cannot 
have $P_1 = 1 - P_2$. 
Hence, since $P_1$ and $P_2$ are orthogonal, we have that 
$P_1 < 1 - P_2$.  
Since $p (P_1 | H ) = 1$ and $p ( (1-P_2 ) | H ) = 0$, this pair
of projections violates the subspace 
implication $P_1 \Rightarrow 1-P_2$. 
That is, a contrary inference implies the existence of   
consistent histories $H$ and $H'$, belonging to different 
consistent sets and agreeing on 
all but one projector, such that $H$ has non-zero
probability, $H'$ has zero probability, and $H < H'$: 
in the example of the last section, for instance, we have
$H = \{ P_{b_+} , P_c \} $ and 
$H' = \{ ( 1 - P_{b_-}) , P_c \}$.  

Clearly, according to the consistent histories 
formalism, an observation of the 
datum $P_{b_+}$ cannot be taken to imply an observation 
of the strictly larger projector $(1 - P_{b_-})$.  To make 
that inference would lead directly to a contradiction, in the 
form of the realisation of 
a probability zero history, if $P_c$ were 
subsequently observed.  

Now it is easy to produce real world examples of contrary inferences, 
so long as those inferences are of unobserved quantities.
As we have seen, it requires only a three-dimensional 
quantum system, prepared in one state, isolated,
and then observed in another state --- hardly a taxing experiment.  

It is not obvious, though, that we can produce examples where
subspace implications fail in a realistic consistent
histories description of {\it observations} of laboratory
experiments, or more generally of macroscopic quasiclassical physics.
That general consistent histories violate subspace implications need
not necessarily imply that the particular consistent histories used 
to recover standard descriptions of real world physics do so.
Both in order to address this question, and for its own sake,
it is interesting to ask whether there might be any alternative 
treatment of quantum theory within the consistent histories
framework which respects subspace implications, at least when 
they relate two consistent histories. 
The next section suggests such a treatment. 

\newsec{Ordered Consistent Sets of Histories} 

We have already seen that there is 
a natural partial ordering for each of the standard 
representations of quantum histories in the consistent histories
approach.  The probability weight defines a second partial ordering:  
$H \prec H'$ if $p(H) < p(H' )$.  
The violation of subspace implications reflects the disagreement 
between these two partial orderings in the consistent histories
formalism: we can have both $H < H'$ and $H \succ H'$.  
The aim of this section is to develop a history-based interpretation 
which restricts attention to collections of quantum histories on which
the two orderings do not disagree.  

We begin at the level of individual quantum
histories, defining an {\it ordered consistent history}, $H$, 
to be a consistent history with the properties that: 
\item{(i)} for all consistent histories $H'$ with $H' \geq H$ we have that 
$p(H' ) \geq p(H)$; 
\item{(ii)} for all consistent histories $H'$ with 
$H' \leq H$ we have that $p(H' ) \leq p(H)$. 

Recall that a consistent history is any quantum history which 
belongs to {\it some} consistent set of histories.  Properties 
(i) and (ii) hold trivially for histories $H$ and $H'$ which belong
to the same consistent set: it is the comparison across different
sets which makes them useful constraints.

We now define an {\it ordered consistent set of histories} to be a 
complete set of exclusive alternative histories, each of which is 
ordered consistent.  We can then define an ordered consistent
histories approach to quantum theory in precise analogy to the 
consistent histories approach, using the same definition of
probability weight and the same interpretation, 
simply declaring by fiat that only ordered consistent sets 
of histories are to be considered.\foot{Note that there are other 
collections of quantum histories on which the orderings do not
disagree.  For example, if all the consistent histories $H$ that 
violate (i) are eliminated, the remainder form a collection on
which the orderings do not disagree and which is not obviously 
identical to the ordered consistent histories, and similarly
for (ii).  It might be interesting to explore such alternatives,
but we restrict attention to the ordered consistent histories
here.}

The following lemmas show that, within the projection operator
formulation, ordered consistent sets of histories
do indeed exist.\foot{I am 
grateful to Bob Griffiths for suggesting a slight extension of Lemma 1.} 
\vskip.1in
{\it Lemma 1:} \qquad  Any consistent history $H = \{ P_1 , \ldots , P_n \}$ 
defined by a series of 
projections which include a minimal projection
$P_j$, so that $P_i \geq P_j$ for all i, is an ordered 
consistent history.\medskip\noindent 
\vskip.1in
{\it Proof:} \qquad  Suppose $H' = \{ P'_1 , \ldots , P'_n \} $ is a 
larger history than $H$, and write $P'_i = P_j + Q_i$. 
We have that 
\eqn\larger{\eqalign{
p ( H' ) & =  \Tr ( P'_n \ldots P'_1 \rho P'_1 \ldots P'_n ) \cr 
         & =  \Tr ( ( P_j + Q_n ) \ldots (P_j + Q_1 ) \rho 
                      (P_j + Q_1 ) \ldots ( P_j + Q_n ) ) \cr
        & =  \Tr ( P_j \rho ) + \Tr (Q_n \ldots Q_1 \rho Q_1 \ldots Q_n ) \cr
        & \geq  \Tr (P_j \rho ) \cr 
        & =  p(H) \, . }}
Now suppose that $H'$ is smaller than H.  Then in particular
$P'_j \leq P_j$ and we have that 
\eqn\smaller{\eqalign{
p(H') & \leq  \Tr (P'_j \rho ) \cr
      & \leq  \Tr (P_j \rho ) \cr
      & = p(H) \, .}}
\vskip .2in 
{\it Lemma 2:} \qquad 
Let $\S = \{ \sigma_1 , \ldots , \sigma_n \}$ be a consistent set of histories 
defined by a series of projective decompositions, of which one, $\sigma_j$,
has the property that for each projection $P$ in $\sigma_j$, and 
for every $i$, we have that there is precisely one 
projection $Q$ in $\sigma_i$ with the property that $Q \geq P$ 
(so that all of the other projections in $\sigma_i$ are contrary to $P$).  
Then $\S$ is an ordered consistent set of histories. 
\vskip.1in
{\it Proof:} \qquad  Each of the histories of non-zero probability in
$\S$ satisfies the conditions of Lemma 1 and so is ordered consistent.
Each of the histories of zero probability in $\S$ is of 
the form $H = \{ \ldots, P, \ldots , Q
, \ldots \}$, where $P$ and $Q$ are contrary projections. 
Now any consistent history smaller than $H$ therefore also contains
a pair of contrary projections $P' \leq P$ and $Q' \leq Q$. 
By the consistency axioms, its probability is less than or equal
to $\Tr (Q'P' \rho P' Q') = 0$, and thus must also be zero.  
Hence $H$ is ordered consistent, 
since any consistent history larger than $H$ has 
probability greater than or equal to zero.  

\newsec{Ordered Consistent Sets and Quasiclassicality} 

A formalism based on ordered consistent sets of histories obviously
defines a more strongly predictive
version of quantum theory than that defined by the existing
consistent histories framework, since it allows strictly
fewer sets as possible descriptions of physics. 
But can it describe our empirical observations?   

The question subdivides.  
Are quasiclassical domains generally ordered consistent sets of histories? 
Is our own quasiclassical domain one? 
If not, are its histories generally ordered when 
compared to histories belonging to other consistent sets 
defined by projections onto ranges of the same 
quasiclassical variables?  For example, can we show that consistent
histories defined by projections onto ranges of densities for chemical
species in small volumes are generally ordered with respect to one
another?  If so, then the type of subspace implication which is generally
used in analyses of observations could still be justified. 
Finally, if either of the previous two properties fail to hold, 
it would be useful to quantify the extent to which they fail.  

Answering any of these questions definitively may 
require --- and, it might be hoped, help to develop --- a deeper
understanding of quasiclassicality than is available to us at
present.  I at any rate do not know the answers, and can 
only offer the questions as interesting ones whose resolution
would have significant implications.  
At least ordered consistency does not seem to fall at the 
first hurdle: ordered consistent sets are shown below to 
be adequate to describe quasiclassicality in simple models. 

As a simple example, consider the following model of a series of 
successive measurements of the spin of a spin-$1/2$ particle 
about various axes.
We use a vector notation for the particle states, 
so that if ${\bf u}$ 
is a unit vector in $R^3$ the eigenstates of $\sigma. {\bf u }$ are
represented by $| \bf \pm u \rangle$.  With the analogy of a pointer
state in mind, we use the basis $\{ |\uparrow\rangle_k ,
|\downarrow\rangle_k \}$ to represent the 
$k^{{\rm th}}$
environment particle state, together with the linear combinations
$|\pm\rangle_k = (|\uparrow\rangle_k \pm |\downarrow\rangle_k
)/\sqrt{2}$.  We compactify the notation by writing environment states
as single kets, so that for example $ |\uparrow\rangle_1 \otimes
\cdots \otimes |\uparrow\rangle_n $ is written as $| \uparrow \ldots
\uparrow \rangle$, and we take the initial state $|\psi(0)\rangle$
to be $|{\bf v}\rangle \otimes | \uparrow \ldots \uparrow
\rangle$.

The interaction between the system and the $k^{{\rm th}}$
environment particle is chosen so that it corresponds to a measurement
of the system spin along the ${\bf u}_k$ direction, so that the states
evolve as follows:
\eqn\measurementinteraction{\eqalign{
  |{\bf u}_k\rangle \otimes |\uparrow\rangle_k & \to  |{\bf
    u}_k\rangle \otimes |\uparrow\rangle_k \, , \cr
  |{\bf -u}_k \rangle
    \otimes |\uparrow\rangle_k & \to  |{\bf -u}_k \rangle \otimes
    |\downarrow\rangle_k \, .}}
A simple unitary operator that generates this evolution is
\eqn\Ukdef{  U_k( t ) = P({\bf u}_k) \otimes I_k + P({\bf -u}_k)
\otimes  \exp (-i\theta_k(t) F_k ) \, ,}
where $P({\bf x}) = |{\bf x}\rangle \langle{\bf x}|$ and $F_k =
i|\downarrow\rangle_k \langle\uparrow|_k - i|\uparrow\rangle_k
\langle\downarrow|_k$.  Here $\theta_k(t)$ is a function defined for
each particle $k$, which varies from $0$ to $\pi/2$ and represents how
far the interaction has progressed.  We define $P_k ({ \pm}) = |{ \pm
}\rangle_k \langle{ \pm}|_k $, so that $F_k = P_k (+)-P_k (-)$.

The Hamiltonian for this interaction is thus
\eqn\hamschrodheis{
  H_{k}(t) = i \dot U_k (t) U_k^\dagger (t)  = \dot \theta_k(t)
  P({\bf -u}_k) \otimes F_k \, ,}
in both the Schr\"odinger and Heisenberg pictures.  We write the
extension of $U_k$ to the total Hilbert space as
\eqn\Vkdef{ 
  V_k = P({\bf u}_k) \otimes I_1 \otimes \cdots \otimes I_n + P({\bf
    -u}_k) \otimes I_1 \otimes \cdots \otimes I_{k-1} \otimes
    \exp ( -i\theta_k(t) F_k ) \otimes I_{k+1} \otimes \cdots
    \otimes I_n \,.}
We take the system particle to interact initially with particle $1$
and then with consecutively numbered ones, and there is no interaction
between environment particles, so that the evolution operator for the
complete system is
\eqn\udefn{  U(t) = V_n(t) \ldots V_1(t) \, ,}
with each factor affecting only the Hilbert spaces of the system and
one of the environment spins.

We suppose, finally, that the interactions take place in disjoint time
intervals and that the first interaction begins at $t=0$, so that the
total Hamiltonian is simply
\eqn\htotal{  H (t ) = \sum_{k=1}^n H_k (t) \, ,}
and we have that $\theta_1 (t) > 0 $ for $t > 0$ and that, if
$0 < \theta_k(t) < \pi/2  $, then $\theta_i(t) = \pi/2
{\rm~for~all~} i < k$ and $\theta_i(t) = 0 {\rm~for~all~} i >k$.

This model has been used elsewhere\refs{\kentmcelwaine,
\mcelwainemaxinfo} in order to explore algorithms 
which might select a single physically natural consistent set 
when the physics is determined by the simplest type of 
system-environment interaction.  
It is particularly well suited to such an analysis, since the dynamics
are chosen so as to allow a simple and quite elegant 
classification\refs{\mcelwainemaxinfo} 
of all the consistent sets built from projections onto subspaces 
defined by the Schmidt decomposition.
Apart from this, though, the model is unexceptional --- one 
of the simpler variants among the many models used
in the literature to investigate the decoherence of system states 
by measurement-type interactions with an environment.   

To give a physical interpretation of the model, we take it  
that the environment ``pointer''
variables assume definite values after their respective 
interactions with the system.  
That is, after the $k^{{\rm th}}$ interaction, 
the $k^{{\rm th}}$ environment particle is in
one of the states $|\uparrow\rangle_k$ 
and $|\downarrow\rangle_k$: the probabilities of each of these
outcomes depend on the outcome of the previous 
measurement (or, in the case of the first measurement, on the
initial state) via the standard quantum mechanical expressions. 

This description can be recovered from the consistent histories
formalism by choosing the consistent set $\S_1$, defined by the 
decompositions
\eqn\projform{ \eqalign{
\{ & I \otimes  | \eps_1 \rangle_1 \langle \eps_1 |_1 \otimes I \otimes \cdots
\otimes I : \eps_1 = \uparrow
{\rm~or~} \downarrow \} {\rm~at~time~} t_1 \, , \cr
\{ & I \otimes  | \eps_1 \rangle_1  \langle \eps_1 |_1 
 \otimes  | \eps_2 \rangle_2  \langle \eps_2 |_2 \otimes \cdots
\otimes I  : \eps_1 , \eps_2 = \uparrow
{\rm~or~} \downarrow \} {\rm~at~time~} t_2 \, , \cr 
& \ldots \cr 
\{ & I \otimes  | \eps_1  \rangle_1  \langle \eps_1 |_1 
 \otimes  | \eps_2 \rangle_2  \langle \eps_2 |_2 \otimes \cdots
 \otimes  | \eps_n \rangle_n  \langle \eps_n |_n 
 : \eps_1 , \eps_2 , \ldots ,
\eps_n = \uparrow
{\rm~or~} \downarrow \} {\rm~at~time~} t_n \, .}}
Clearly, the histories of non-zero probability in $\S_1$ take the form
\eqn\seqform{\eqalign{
H_{\eps_1 , \ldots , \eps_n }  = 
\{ & I \otimes I \otimes I \otimes \cdots \otimes I , \cr 
& I \otimes     | \eps_1 \rangle_1 \langle \eps_1 |_1 \otimes I \otimes 
                \cdots \otimes I  , \cr 
& I \otimes     | \eps_1 \rangle_1 \langle \eps_1 |_1 \otimes
| \eps_2 \rangle_2 \langle \eps_2 |_2 \otimes I \otimes \cdots \otimes I ,
\cr 
& \ldots, \cr 
& I \otimes     | \eps_1 \rangle_1 \langle \eps_1 |_1 \otimes
| \eps_2 \rangle_2 \langle \eps_2 |_2 \otimes \cdots \otimes 
| \eps_n \rangle_n \langle \eps_n |_n  \} \, , }}
for sequences $\{ \eps_1 , \ldots , \eps_n \}$, each element of 
which takes the value $\uparrow$ or $\downarrow$. 
Their probabilities, defined by the decoherence functional,
are precisely those which would be obtained from standard quantum 
theory by treating each interaction as a measurement: 
\eqn\probs{
p ( H_{\eps_1 , \ldots , \eps_n } ) = 
( {{ 1 + a_1 {\bf v . u_1 } } \over {2}} ) 
( {{ 1 + a_2 {\bf u_1 . u_2 } } \over {2}} ) 
\ldots 
( {{ 1 + a_n {\bf u_{n-1} . u_n } } \over { 2}} ) \, , }
where, letting $\eps_0 = \uparrow$, we define
$a_i = 1$ if $\eps_i$ and $\eps_{i-1}$ take the same value, 
and $a_i = -1$ otherwise. 

Now $\S_1$ is defined by a nested sequence of increasingly 
refined projective decompositions, all of whose
projections commute --- a relation which is unaltered by 
moving to the Heisenberg picture.  
It therefore satisfies the conditions of Lemma 2 above, and
so is ordered consistent.  

This argument clearly generalizes: in any situation in which 
Hilbert space factorizes into system and environment degrees of
freedom, where the self-interactions of the latter are negligible,
any consistent set defined by nested commuting projections onto
the environment variables is ordered consistent. 
The model considered above is a particularly 
crude example: more sophisticated, and phenomenologically somewhat more 
plausible, examples of this type are analysed in, for example,
Refs. \refs{\gmhstrongd, \gmhprd}.  

No sweeping conclusion can be drawn from this, 
since it is generally agreed that familiar quasiclassical physics  
is {\it not} well described in general --- at least in 
any obvious way --- by models of this type.  (Again, a detailed discussion 
of the limitations of such models can be found in Refs. 
\refs{\gmhstrongd, \gmhprd}.) 
In other words, while it would be 
hard to defend the hypothesis that familiar quasiclassical sets are
generally ordered consistent if sets of the type $\S_1$ were 
not, the fact that they are is certainly not sufficient evidence. 
It would be good to find sharper tests of the 
hypothesis, perhaps for example by developing further the phenomenological
investigations of quasiclassicality pursued in Ref. \refs{\gmhstrongd}.  
Meanwhile, the questions raised earlier in this section remain 
unresolved.   

On the other hand, it would be difficult to make a watertight case that 
ordered consistent sets are definitely inadequate 
to describe real-world physics, for the following reason.  
First, it seems hard to exclude the possibility that the initial
state is pure, so let us temporarily suppose that it 
is: $\rho = | \psi \rangle \langle \psi | $. 
As Gell-Mann and Hartle point out\refs{\gmhprd}, 
we can then associate to every consistent set of histories, $\S$, 
a nested set of commuting projections defining
what they term generalized records. 
The consistent set defines a resolution of the initial state into
history vectors, 
\eqn\resolution{
| \psi \rangle = \sum_{i_1 , \ldots , i_n } P^{i_n}_n \ldots P^{i_1}_1
| \psi \rangle  \, ,}
which are guaranteed to be orthogonal by the consistency condition
\decohgmha . 
We can thus find at least one set of orthogonal projection operators 
$\{ R_I \}$, indexed by sets of the form $I = \{ i_1 \ldots i_n \}$,
which project onto the history vectors 
and sum to the identity:
\eqn\rconds{\eqalign{
R_{i_1 \ldots i_n } | \psi \rangle & = P^{i_n}_n \ldots P^{i_1}_1
| \psi \rangle \, , \cr
\sum_I R_I & = I \, , \cr 
R_I R_J & = \delta_{IJ} R_J \, .}}
We can thus\refs{\gmhprd} construct a set $\S'$, with the same history vectors
and the same probabilities as $\S$, built from a nested sequence
of commuting projections defined by sums of the $R_I$.  
And, as we have seen, sets of this type are ordered consistent. 

There is no reason to expect the 
projections defining the set $\S'$ to be closely related to
those defining $\S$.  In particular, the fact that $\S$ is 
a quasiclassical domain certainly does not imply that 
$\S'$ is likely to be: its projections are not generally likely to
be interpretable in terms of familiar variables.  
But, as we have already noted, we have no theoretical criterion
which identifies a particular consistent set, or a particular
type of variable, as fundamentally correct for representing the
events we observe.  The set $\S'$ correctly identifies the history
vectors and predicts their probabilities, and we thus could not say for
certain (given that we presently have no theory of set selection)
that its description of physics 
is fundamentally incorrect, while that given by $\S$ 
is fundamentally correct.  

To make this observation is merely to point out a 
logical possibility.  
In fact, it would be extremely puzzling
if the more complicated and apparently derivative set $\S'$ 
were in some sense more fundamental than the associated
quasiclassical domain $\S$.  And even if this were 
somehow understood to be true in principle, we would still need
to understand the relationship between between ordered consistency and 
quasiclassicality in order to say whether or not standardly used 
subspace inferences are in fact justifiable. 

\newsec{Ordering and Ordering Violations: Interpretation} 

However, one of the main points of this
paper --- and the main reason for taking a particular interest in 
the properties of ordered consistent sets --- is that either answer 
leads to an interesting conclusion.  

If our empirical
observations can be accounted for by the predictions of an 
ordered consistent set, then the ordered consistent sets formalism
supersedes the current versions of the consistent histories formalism 
as a predictive theory. 
Alternatively, if the predictions of ordered consistent histories 
quantum theory are false, then either the consistent histories 
framework uses entirely the wrong language to describe histories 
of events in quantum theory, or we cannot generally rely on  
subspace implications in analysing our observations.  

Either of these last two possibilities would have far-reaching 
implications for our understanding of nature.  It is true that
other ways of representing quantum histories are known than those used 
in the consistent histories formalism, but they arise either in
non-standard versions of quantum theory, such as de Broglie--Bohm
theory, or in alternative theories. 
It is also true that there is no way of logically excluding the 
possibility that subspace implications generally fail to hold.
Any clear violation would, however, lead to radical changes in
our representation of the world, and in particular to our 
understanding of the relation between theory and empirical observation.  

I would 
suggest, however, that any 
version of closed system quantum theory in which the two 
orderings disagree leads to radical new interpretational problems.  
The fundamental problem is that, supposing that the world we
experience is described by one particular realised quantum
history, we never know --- no matter how precise we try to make our
observations --- exactly what form that history takes.    
This is not only because we can never completely eliminate imprecision
from our experimental observations.  
A deeper problem is that we have no theoretical understanding 
of how, precisely, an observation should be represented 
within quantum theory.  We do not know precisely 
when and where any given observation takes place.
Nor do we know whether is fundamentally correct to
represent quantum events by projection operators, by quantum effects, 
by statements associated to space-time regions in path integral
quantum theory, or in some other way --- let alone precisely 
which operator, effect, or statement correctly represents any given event.  
As a result, we are always forced
into guesswork and approximation.  We are forced to assume, at least as a
working hypothesis, that we can find sensible bounds on 
our observations. 
Roughly speaking, we assume that we can say, at least, that a photon
hit our photographic plate within a certain region, that the observed 
flux from a distant star was in a certain range, and so 
forth.\foot{In fact such statements are generally made
within statistical confidence limits.  To consider statistical statements
would complicate the discussion a little, but does not alter the 
underlying point.}  We assume also that the 
probability of the actual observations --- whose precise form
we do not know --- is bounded by the probability of the observations
as we approximately represent them.  These assumptions ultimately rely on 
the agreement of two orderings just mentioned: when those orderings 
disagree, we therefore run into new problems.  

It is easy to see, in particular, that this sort of problem arises 
in any careful consistent histories treatment of quantum cosmology.  
Suppose, for example, that we have a sequence of cosmological 
events which we wish to represent theoretically, in order to 
calculate their probability, given some theory of the boundary
conditions.  Assuming that the basic principles of the
consistent histories formalism are correct, we know 
that these events should be represented by 
some history $H$ belonging to some consistent set $\S$. 
We do not, however, know the precise form of $H$ or of $\S$: 
the events are given to us as empirical observations rather than
as mathematical constructs.  

The best we can then do, following the general
principles of the consistent histories formalism, is choose some
plausible consistent set $\S'$ containing histories $H_{\rm min}$ 
and $H_{\rm max}$ which we guess to have the property 
$H_{\rm min} < H < H_{\rm max}$: in particular, thus, we choose
$H_{\rm min} < H_{\rm max}$.  Since $H_{\rm min}$ and $H_{\rm max}$
belong to the same set, we have that $p(H_{\rm min}) < p(H_{\rm
max})$.  It might naively be hoped that we can derive that 
$p(H_{\rm min}) < p(H) < p(H_{\rm max})$, but since $H$ in 
general will belong to a different 
consistent set from  $H_{\rm min}$ and $H_{\rm max}$, this does
not generally follow.  
There is no way to bound $p(H)$, except (in 
principle) by performing the enormous task of explicitly 
calculating the probabilities of {\it all} 
consistent histories bounded by $H_{\rm min}$ and $H_{\rm max}$, 
and there is no way to justify 
the type of subspace implication --- relating observations and 
true data --- that we generally take for granted.  

This is not to say that the disagreement of the two orderings 
necessarily leads to logical contradiction.  Versions of quantum theory in 
which the orderings disagree need not be inconsistent, or even
impossible to test precisely.
They do, though, generally seem to require us to identify {\it
precisely} the correct representation of our observations in 
quantum theory.  This is generally a far from trivial 
problem: how are we to tell, a priori, exactly which projection
operators represent the results of a series of quantum measurements? 
It is not impossible to imagine that theoretical criteria could
be found which solve the problem, but we certainly do not have such 
criteria at present. 

\vskip .2in 
\newsec{Conclusions}

Though the criterion of ordering seems mathematically natural, both 
in the consistent histories approach to quantum theory and
in other possible treatments of quantum histories, it raises 
very unconventional questions.  
It seems, though, that these questions cannot be avoided
in any precise formulation of the quantum theory of a closed 
system which involves a standard representation of quantum
events and which gives a historical account. 

There seem to be three possibilities, each of which is interesting.
The first is that 
the representations of quantum histories discussed here, 
though standard, are not those chosen by nature.  
Clearly this is a possibility: there are, for example, well known 
non-standard versions of quantum theory\refs{\bohm}, and 
related theories\refs{\grw}, in which histories are defined by 
trajectories or other auxiliary variables, and in which 
subspace implications follow just as in classical physics.  

The second possibility is that our quasiclassical domain can be
shown to be an ordered consistent set. If so, then 
the ordered consistent histories approach is both predictively
stronger than the standard consistent histories approach ---
since there are fewer ordered consistent sets --- and 
compatible with empirical observation, and hence superior
as a predictive theory. 
If it is compatible with our observations, the ordered consistent
histories approach would seem at least as natural as the consistent
histories approach.  

Even if so, I would not  
suggest that the ordered consistent histories formalism is 
the ``right'' interpretation of quantum theory, and 
the consistent histories approach the ``wrong'' one.
The ordered consistent histories approach seems 
almost certain to suffer from many of the same defects as
the consistent histories approach, since there are still far
too many ordered consistent sets.  
The aim here is thus not to propose the ordered consistent
histories approach as a plausible fundamental interpretation
of quantum theory, but to suggest that the range of natural and 
useful mathematical definitions of 
types of quantum history is wider than previously understood. 
This range includes, at least, Goldstein and Page's criterion of linear 
positivity\refs{\goldsteinpage}, the various
consistency criteria\refs{\griff,\gmhstrongd,\gmhprd} in the 
literature, and the criterion of ordered consistency introduced here. 
It seems to me hard to justify taking any of 
these criteria as defining the fundamentally correct interpretation of 
quantum theory.  On the one hand, physically interesting quantum histories 
might possibly satisfy any one, or none, of them; on the other hand,
most quantum histories satisfying any given criterion seem unlikely
to be physically interesting --- and precisely which criteria are useful 
in which circumstances largely remains to be 
understood.  

The third possibility is that our quasiclassical domain is not 
an ordered consistent set. This would have intriguing theoretical 
implications.  We would have, at least in principle, to abandon 
subspace inferences, and we would ultimately
need to understand precisely how to characterise the 
quantum events which constitute the history we observe.
This would raise profound and not easily answerable 
questions about how we can tell what, precisely, {\it are} 
our empirical observations. 
It might also, depending on the way in which ordered consistency
was violated, and the extent of any violation, raise 
significant practical problems in the analysis of 
those observations. 

No compelling argument in favour of any one
of these possibilities has been given here: it has been shown 
only that, if quasiclassical sets generally fail to be ordered consistent, 
they do so in a way too subtle to be displayed in the simplest 
models.  

Another caveat is that the above discussion applied the 
criterion of ordered consistency only to the simplest
representation of quantum histories, in which individual
events are represented by projections at a single time. 
Other representations need to be considered case by case,
and our conclusions might not necessarily generalize.  
For example, the fact that a consistent history built from
single time projections is ordered when compared to 
consistent histories of the same type does not necessarily 
imply that it is ordered when compared to consistent histories
defined by composite events. 

Still, the criterion of ordered consistency defines a new
version of the consistent histories formulation of quantum 
theory, which avoids the 
problems caused by contrary inferences.  Its other properties and 
implications largely remain to be understood. 
\vskip1in
\leftline{\bf Acknowledgements}  
I am grateful to Jeremy Butterfield 
for a critical reading of the manuscript and many thoughtful comments,
to Chris Isham and Noah Linden for very helpful discussions of
their related work, and to Fay Dowker, Arthur Fine and Bob 
Griffiths for helpful comments.

I would particularly like to thank Francesco Petruccione for 
organising the small and lively meeting at which this work,
inter alia, was discussed, and for his patient editorial
encouragement. 

This work was supported by a Royal Society University 
Research Fellowship. 
\listrefs

\leftline{\bf Appendix: Ordering and Decoherence Functionals}

This Appendix describes a noteworthy earlier discussion of quantum 
history orderings, given by Isham and Linden in Sec. IV 
of Ref. \refs{\ILqtl}, and its relation to the ideas
discussed here. 

Isham and Linden abstract the basic ideas of the consistent histories
formalism in the following way.  First, the space ${\cal UP}$ of {\it history
propositions} is taken to be a mathematical structure --- 
an orthoalgebra --- with a series of operations and
relations obeying certain axioms.  In particular, they propose that a 
partial ordering $\leq$ and an orthogonality relation $\perp$ should 
be defined on ${\cal UP}$ and should obey natural rules, and that 
${\cal UP}$ should include an identity history $1$.  

They then introduce a space ${\cal D}$ of decoherence functionals, 
defined to be maps from ${\cal UP} \times {\cal UP}$ to the complex numbers
satisfying certain axioms, and go on to consider whether the 
axioms defining decoherence functionals should include axioms
relating to the ordering in ${\cal UP}$.  

In the language of standard quantum mechanics, ${\cal UP}$ corresponds
to the space of all the quantum histories (not only the consistent histories)
for a given system, whose Hilbert space and hamiltonian are fixed.    
Any of several representations of quantum histories could be considered: 
the relevant part of Isham and Linden's discussion uses the simplest
representation of quantum histories, as sequences of projection
operators.  

The standard quantum mechanical decoherence functional (as
appears on the left hand side of \decohgmha\ ) is a member of 
the space ${\cal D}$ in the minimal axiom system Isham 
and Linden eventually choose.  As they remark, though, 
it would not be a member of ${\cal D}$ if the extra ordering axioms
they discuss were imposed.  
Isham and Linden nonetheless consider imposing these
ordering axioms, since their aim in the relevant discussion is to 
investigate generalised algebraic and logical schemes rather than
to propose a formalism applicable to standard quantum 
theory.  (They suggest, at the end of section IV, that standard quantum
theory might perhaps emerge from some such generalised scheme in an
appropriate limit.)

Isham and Linden were, as far as I am aware, the first to investigate
possible uses of orderings in developing the consistent histories
formalism.  
It is worth stressing, though, to avoid 
any possible confusion, that their suggestions pursue the 
exploration of orderings in a direction orthogonal to the one 
considered in the present paper.  
In this paper we restrict attention to standard quantum theory, and 
propose an alternative histories formalism within that theory, 
using the standard quantum theoretic decoherence functional 
throughout.  We note also that the subspace implications which
underlie our basic scientific worldviews depend for their 
justification on the assumption that the quasiclassical set 
describing the physics we observe is an ordered consistent set. 
Isham and Linden's proposed ordering axioms, on the other hand, 
exclude standard quantum theory and the standard decoherence
functional: they are possible postulates which might be imposed
on non-standard generalised decoherence functionals in non-standard
generalisations of quantum theory. 

Isham and Linden give a minimal set of postulated properties for 
generalised decoherence functionals: 
\eqn\minposts{\eqalign{
d(0,\a) & =0 \, {\rm~for~all~} \a\in\UP \, ; \cr
d(\a,\b)& =d(\b,\a)^* \, {\rm~for~all~} \a,\b\in\UP \, ; \cr
d(\a,\a)& \ge 0 \, {\rm~for~all~} \a\in\UP \, ; \cr 
{\rm~if~} \a\perp & \, \b  \, {\rm~ then}, {\rm~for~all~} \g \, , \,
d(\a\oplus\b,\g)=d(\a,\g)+d(\b,\g) \, ; \cr 
d(1,1) & =1 \, .}
}

They then consider imposing new postulates on decoherence
functionals.  The first of these --- their posited inequality 1 --- 
is that: 
\eqn\posineqone{
{\rm for\ all\ } d\in\D {\rm\ and\ for\ all\ }\a,\b {\rm\ with\ }
     \a\leq\b {\rm\ we\ have\ } d(\a,\a)\leq d(\b,\b) \, .}

As Isham and Linden go on to point out, there are  
familiar examples in standard quantum theory in which \posineqone\ is 
violated for a pair of histories $\a \leq \b$ in which one of the
histories (in their case $\a$) is inconsistent.\foot{The 
suggestion that inequality 1 is true when applied 
to sequences of projectors 
onto subsets of configuration space in a path-integral quantum 
theory is thus misleading: it is easy to find configuration
space analogues of these examples.  I am grateful to Chris Isham 
and Noah Linden for discussions of this point.}

Two further postulates on generalised decoherence 
functionals are also posited: 
\eqn\posineqtwo{
    \a\perp\b{\rm\ implies\ }
        d(\a,\a)+d(\b,\b)\leq 1{\rm\ for\ all\ }d\in\D \, ;}
and 
\eqn\posineqthree{
        {\rm for\ all\ } d\in\D{\rm\ and\ all\ } \g\in\UP
                {\rm\ we\ have\ }d(\g,\g)\leq 1 \, .}
Isham and Linden give examples to show that, in standard 
quantum theory, with the
standard decoherence functional, inconsistent histories do not
necessarily respect these inequalities either.   

Again, the difference from the examples considered in the present
paper is worth emphasizing. 
All of the examples Isham and Linden consider involve inconsistent
histories --- these are all they require 
in order to investigate possible properties of decoherence 
functionals applied to arbitrary, not necessarily 
consistent, quantum histories.  
These examples are not problematic for the consistent 
histories approach to quantum theory, according to which 
the inconsistent histories have no physical significance, and 
they do not give rise to new interpretational questions in  
any conventional quantum histories approach, for essentially the same reason. 
The discussion in the present paper, on the other hand, looks at the 
properties of {\it consistent} histories in standard quantum theory: 
we have argued that their failure to respect ordering relations is 
problematic and explained that it does raise new questions. 

Suppose now that we set aside Isham and Linden's 
motivations, and alter their ordering postulates so that they 
apply, not to generalised decoherence functionals applied to all quantum
histories in an abstract generalisation of quantum theory, but to 
the standard decoherence functional applied to ordered consistent 
histories in standard quantum theory. 
We then obtain the following:
\eqn\posordone{ {\rm for\ all\ }\a,\b {\rm\ with\ }
     \a\leq\b {\rm\ we\ have\ } d(\a,\a)\leq d(\b,\b) \, ;}
\eqn\posordqtwo{
    \a\perp\b{\rm\ implies\ }
        d(\a,\a)+d(\b,\b)\leq 1 \, ;}
and 
\eqn\posineqthree{
        {\rm for\ all\ } \g 
                {\rm\ we\ have\ }d(\g,\g)\leq 1 \, .}
Here $d$ is the standard decoherence functional,
$\a$, $\b$ and $\g$ are now taken to be ordered consistent histories,
and $\a \perp \b$ means that $\a$ and $\b$ are disjoint --- i.e.,
there is at least one time at which their respective events are
represented by contrary projections.  

The first of these equations holds by the definition of an ordered consistent
history, but it might perhaps be hoped that the others could 
restrict the class of histories further.
However, the second equation also holds for all ordered consistent histories.
To see this, note that $\a\perp\b$ implies that $\b \leq (1-\a)$, 
and that if $\a$ is a consistent history then $(1- \a )$ is too. 
The fact that $\b$ is ordered consistent thus implies that 
\eqn\boc{p( \b ) \leq p ( 1 - \a ) = 1 - p (\a ) \, .}
The third equation, moreover, holds for all consistent histories, ordered or
otherwise.   It seems, then, that ordered consistency may be the 
strongest natural criterion that can be defined using the 
basic ingredients of consistency and ordering.

\end